\documentclass{amia}
\pdfoutput=1
\usepackage{graphicx}
\usepackage[labelfont=bf]{caption}
\usepackage[superscript,nomove]{cite}
\usepackage[super,sort&compress]{natbib}
\setcitestyle{citesep={,}}
\usepackage{color}

\begin{document}

\title{An Informatics-based Approach to Identify Key Pharmacological Components in Drug-Drug Interactions}

\author{Jianyuan Deng, M.Phil.$^{1}$, Fusheng Wang, Ph.D.$^{1,2}$}

\institutes{
   $^1$Department of Biomedical Informatics, Stony Brook University\\
   $^2$Department of Computer Science, Stony Brook University
}

\maketitle

\noindent{\bf Abstract}

\textit{Drug-drug interactions (DDI) can cause severe adverse drug reactions and pose a major challenge to medication therapy. Recently, informatics-based approaches are emerging for DDI studies. In this paper, we aim to identify key pharmacological components in DDI based on large-scale data from DrugBank, a comprehensive DDI database. With pharmacological components as features, logistic regression is used to perform DDI classification with a focus on searching for most predictive features, a process of identifying key pharmacological components. Using univariate feature selection with chi-squared statistic as the ranking criteria, our study reveals that top 10\% features can achieve comparable classification performance compared to that using all features. The top 10\% features are identified to be key pharmacological components. 
Furthermore, their importance is quantified by feature coefficients in the classifier, which measures the DDI potential and provides a novel perspective to evaluate pharmacological components. 
} 

\section*{1. Introduction}
Drug-Drug Interactions (DDI) can account for 30\% of Adverse Drug Reactions (ADRs) \cite{tatonetti2011novel}, which lead to 1.3 million emergency department visits and 350,000 hospitalizations in the United States each year \cite{shehab2016us}. Recently, there is an increasing trend of polypharmacy \cite{maher2014clinical}, especially among the elderly people. According to the Centers for Disease Control and Prevention, the percentage of the US population taking five or more prescription drugs in past 30 days increased from 4.0\% during 1988-1994 to 10.9\% during 2011-2014. For seniors 65 years or older, the proportion increased from 13.8\% to 40.7\% \cite{national2018health}, which boosts the incidence of DDIs. Thus, identification of possible DDI before a drug is launched into market becomes critical. 

DDI can be broadly classified into two categories: pharmacokinetic (PK) and pharmacodynamic (PD) DDI. PK DDIs are the cases when a drug affects the PK processes, namely, absorption, distribution, metabolism and excretion (ADME) of another co-administered drug. This will lead to concentration variations of the parent drug or active metabolite at the site of action. PD DDI would happen if one drug has an antagonistic, additive, synergistic or even indirect effect on the action of another drug. Key pharmacological components involved in PK and PD DDI include four groups: carrier, transporter, enzyme and target \cite{ferdousi2017computational}. A carrier is a secreted protein which can bind drugs and carry them to move around the biological fluids. A transporter is a membrane protein which can facilitate in the influx or efflux of xenobiotics. An enzyme is engaged in the bio-transformation of many substances. A target is a biological component which drugs can interact with to exert a direct pharmacodynamic effect. Carriers, transporters and enzymes are usually involved in PK DDI while targets mainly mediate PD DDI.

Traditional strategies to identify DDI are often based on a series of \textit{in vitro} and \textit{in vivo} studies. When the \textit{in vitro} experiment results indicate that the compound interacts with certain pharmacological components, i.e., carriers, transporters, enzymes and targets, subsequent \textit{in vivo} experiments would be conducted to verify the interaction \cite{huang2007drug}. To accelerate wet-lab DDI identification efficiency, \textit{in silico} approaches can also be used, such as 1) predicting whether a compound interacts with certain carrier, target, enzyme or transporter using \textit{in silico} virtual screening approaches \cite{karlgren2012vitro} and 2) studying mechanisms of a compound in ADME using various PKPD models \cite{lehr2010semi, varma2013mechanistic,zhu2016agent, deng2018silico, marsousi2018prediction}. However, most studies only test a limited number of drugs at a time involving several selected pharmacological components, which are considered important in DDI based on previous experience. However, one essential question remains unanswered: which pharmacological components are most relevant to DDI and should be recruited in the follow-up studies, especially considering the high cost of studies and long duration of experiments?

In the past decade, with the rise of `Big Data', informatics-based DDI studies exploiting large-scale data are emerging.  These studies can be classified into two broad categories based on the goals: 1) DDI detection studies and 2) DDI prediction studies.  DDI detection studies focus on detecting novel DDI signals for existing drugs, such as mining into FDA Adverse Event Reporting System (FAERS), social media, literature, and electronic health records (EHR) for potential DDI \cite{tatonetti2011novel, boyce2012using, pathak2013using, yang2015mining,kolchinsky2015extraction, lorberbaum2016integrative, lim2018drug, sun2019drug}.  DDI prediction studies focus on predicting novel DDI signals for new drugs or new drug combinations using drug knowledge databases \cite{tari2010discovering,gottlieb2012indi, vilar2012drug, guimera2013network, huang2013systematic,cheng2014machine,zhang2015label, sridhar2016probabilistic,fokoue2016predicting,ferdousi2017computational,liu2017analysis, zhang2017predicting, kastrin2018predicting, ryu2018deep}. Nonetheless, these studies are not aimed at understanding the pharmacological components through which a DDI occurs, and the question above still remains unanswered. 

In this study, we aim to identify important factors contributing to DDI, which is a feature selection problem \textit{per se} in the context of DDI classification. To achieve this goal, we conduct a DDI classification task with pharmacological components as features, and select most contributing features as the most influential pharmacological components. 

\section*{2. Methods}
\begin{figure}[h!]
\centering
\includegraphics[scale=0.5]{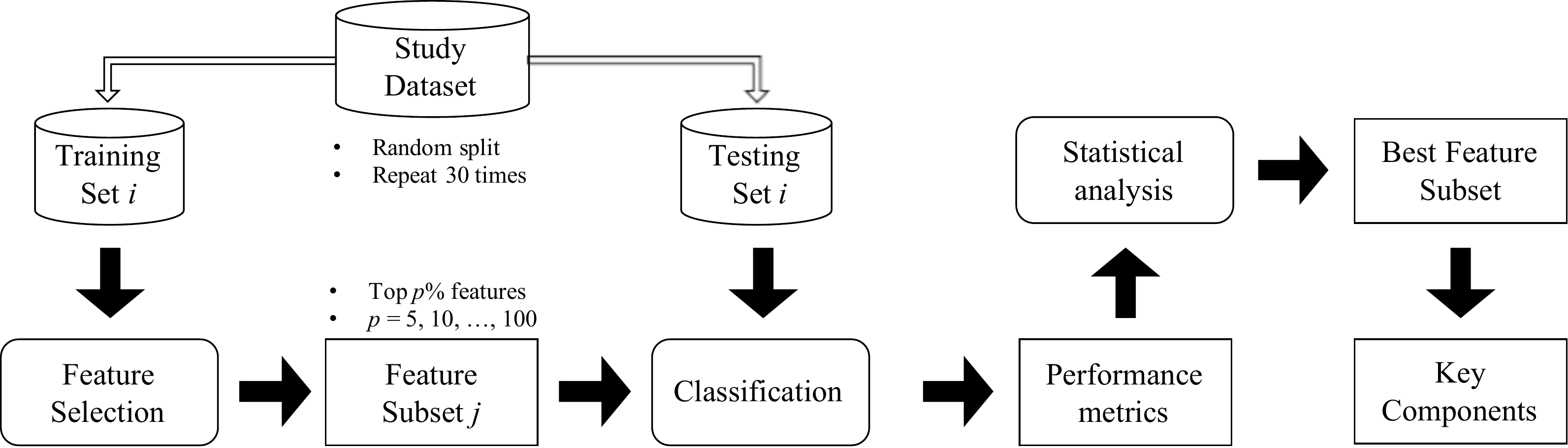}
\caption{Methodology for Key Pharmacological Components Identification}
\label{Methodology}
\end{figure}

The experimental setup is illustrated in Figure ~\ref{Methodology}. We embed the key pharmacological components identification in a typical classification task and search for the most relevant features. We first generate different feature subsets by keeping different proportions of all features in the training set  using univariate feature selection method. We then use these feature subsets on the testing set to perform DDI classification respectively. Finally, we evaluate classification performances of these feature subsets for best feature subset selection. To make the identified pharmacological components more reliable, this process is repeated 30 times.

\subsection*{2.1 DDI Classification}
\textit{2.1.1 DDI Dataset Construction} 

DrugBank is a comprehensive drug database with basic drug information such as pharmacological components that a drug interacts with, as well as a list of DDIs, which originate from drug package inserts and are usually treated as the gold-standard, positive DDI \cite{wishart2017drugbank}. In this study, we use the latest version 5.1.4 (released on 2019-07-02) of DrugBank as the data source. 

\begin{figure}[h!]
\centering
\includegraphics[scale=0.5]{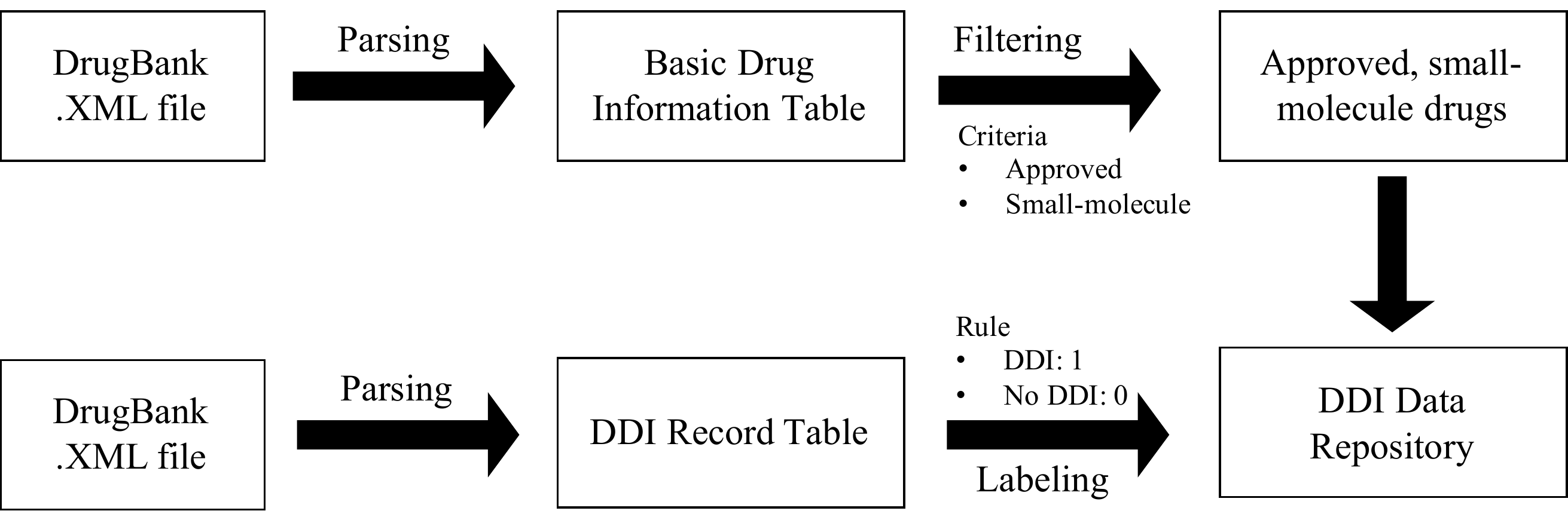}
\caption{Construction of the DDI Data Repository}
\label{DDIRepositoryConstruction}
\end{figure}

We first construct a drug information table with the following information:  1) drug types: "small-molecule" or "biotech", 2) drug groups: "experimental", "investigational" or "approved", and 3) pharmacological components interacting with a drug: carriers, transporters, enzymes and targets. After parsing the DrugBank database in XML format, we get 13,339 drug entries in total. We then construct a DDI table by extracting all the DDI records for each drug, which results in 2,723,944 DDI records. In the final step of constructing DDI data repository, we narrow the study scope to the 2,594 approved, small-molecule drugs, which form 3,363,121 unique drug-drug pairs. A drug-drug pair is an instance in the DDI repository. Next, by querying the basic drug information table, we retrieve the pharmacological components for each drug in all drug-drug pairs. By querying the DDI table, we label DDI status for each instance. If the drug-drug pair is in the DDI table, then an interaction exists and the instance is labeled as a positive DDI case. Otherwise, it is labeled as a negative DDI case. Among the 3,363,121 instances, there are 615,537 positive DDI cases and the positive rate is 18.3\%. The DDI data repository construction process is illustrated in Figure ~\ref{DDIRepositoryConstruction}. 

We randomly pick 200,000 DDI cases from the repository as the current DDI study dataset, with a positive rate of 18.3\%. 

\textit{2.1.2 Feature Engineering}

Since our study objective is to identify key pharmacological components, we use carrier(s), target(s), enzyme(s) and transporter(s) interacting with the two drugs in each DDI instance as features. In the study dataset, a DDI instance $I\textsubscript{i}$ consists of two drugs $D\textsubscript{i, 1}$ and $D\textsubscript{i, 2}$. For drug $D\textsubscript{i, 1}$, the interacting carriers, transporters, enzymes and targets are denoted as $carrier\textsubscript{i, 1}$, $transporter\textsubscript{i, 1}$, $enzyme\textsubscript{i, 1}$ and $target\textsubscript{i, 1}$, respectively. Similarly, for drug $D\textsubscript{i, 2}$, its corresponding pharmacological components are denoted as $carrier\textsubscript{i, 2}$, $transporter\textsubscript{i, 2}$, $enzyme\textsubscript{i, 2}$ and $target\textsubscript{i, 2}$, respectively.

\begin{figure}[h!]
\centering
\includegraphics[scale=0.6]{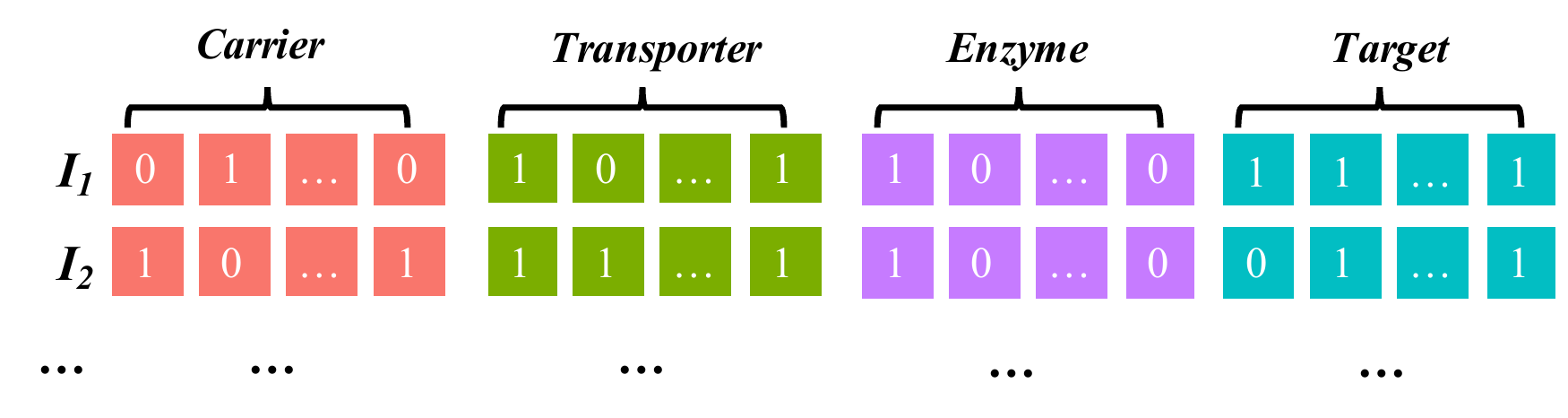}
\caption{One-Hot Encoding for Pharmacological Components}
\label{FeatureEngineering}
\end{figure}

One common hypothesis for the cause of a PK or PD DDI is that there are some overlaps in certain pharmacological components which two drugs can both interact with \cite{ferdousi2017computational}. Thus, we only retain the overlapping pharmacological components for $D\textsubscript{i, 1}$ and $D\textsubscript{i, 2}$ in a given DDI instance $I\textsubscript{i}$ to get  $carrier\textsubscript{i}$, $transporter\textsubscript{i}$, $enzyme\textsubscript{i}$ and $target\textsubscript{i}$. They are then aggregated over all DDI instances to form $Carrier$, $Transporter$, $Enzyme$ and $Target$ and are converted into binary vectors using one-hot encoding as features to classify DDI (Figure ~\ref{FeatureEngineering}). In total, there are 498 features, including 16 carriers,  65 transporters, 62 enzymes and 355 targets.

\textit{2.1.3 Classifier}

In this study, we adopt logistic regression (LR) as the classifier because of its high interpretability. LR uses a logistic function to model a binary target variable $Y$ with features $X\textsubscript{1}, \cdots, X\textsubscript{D}$. The log odds of the event that $Y=1$ is defined in Eq.~\ref{log odds}:
\begin{equation}
    \log\textsubscript{e} \frac{P(Y=1)}{1 - P(Y=1)} = \beta \textsubscript{0} + \beta \textsubscript{1}X\textsubscript{1} + \cdots + \beta \textsubscript{D}X\textsubscript{D} \label{log odds}
\end{equation}
where $\beta \textsubscript{1}, \cdots, \beta \textsubscript{D}$ are coefficients for features $X\textsubscript{1}, \cdots, X\textsubscript{D}$. These coefficients can be interpreted in a straightforward way. For example, if $\beta \textsubscript{1} = 1$, then increasing $X\textsubscript{1}$ from 0 to 1 increases the odds of event $Y=1$ by a factor $e$. In our setting, for a given pharmacological component with $\beta$ coefficient being 1, if two drugs can both interact with the component, the interpretation is that the odds of a positive DDI is increased by $e$ times. 

To counteract the effect of imbalanced classes in our dataset, we adjust the weights to make them inversely proportional to class frequencies. Other hyper parameters in the LR classifier are set based on 5-fold cross validation using the entire study dataset, which consists of 200,000 DDI cases and 498 features. All the hyper parameters are kept constant during subsequent experiments. 

\subsection*{2.2 Feature Selection}
The study datasets are randomly splitted into a training set (50\%) and a testing set (50\%). Univariate feature selection is used to rank all features in the training set in a supervised way. Since the target variable and the features are all binary vectors, we choose chi squared statistic as the ranking criteria. We thereby generate different feature subsets by varying the top percentile of the ranked features, \textit{i.e.}, the proportion of features to be recruited in the classification task. For example, with the whole feature set denoted as $F$, feature subset $F\textsubscript{P}$ represents the top $P\%$ features in $F$ ranked by the chi squared statistic $(P\in \{5,10,15, \cdots, 100\})$. 

\subsection*{2.3 Feature Subset Selection}
Each $F\textsubscript{P}$ selected in the training set is applied in the DDI classification task. Classification performance is examined by the following metrics, namely, accuracy, recall, precision, F1 socre, area under the Receiver Operating Curve (AUROC) and area under the Precision Recall Curve (AUPRC). The classification performance reflects DDI predictive ability of each $F\textsubscript{P}$. The best feature subset $F\textsubscript{P}$ is expected to have the best classification performance. Note that the DDI classification task is repeated 30 times by randomly splitting the study dataset 30 times to appeal to the central limit theorem for later statistical analyses. 

\subsection*{2.4 Identification of Key Features}
With the top percentile $P$ identified, we extract all the features in feature subset $F\textsubscript{P}$ over 30 repeats. For each individual feature, \textit{i.e.}, a pharmacological component, we also extract their corresponding $\beta$ coefficients in the LR classifier as a measure for DDI potential.


\section*{3. Results}
\subsection*{3.1 Best Feature Subset}
\begin{figure}[h!]
\centering
\includegraphics[scale=0.62]{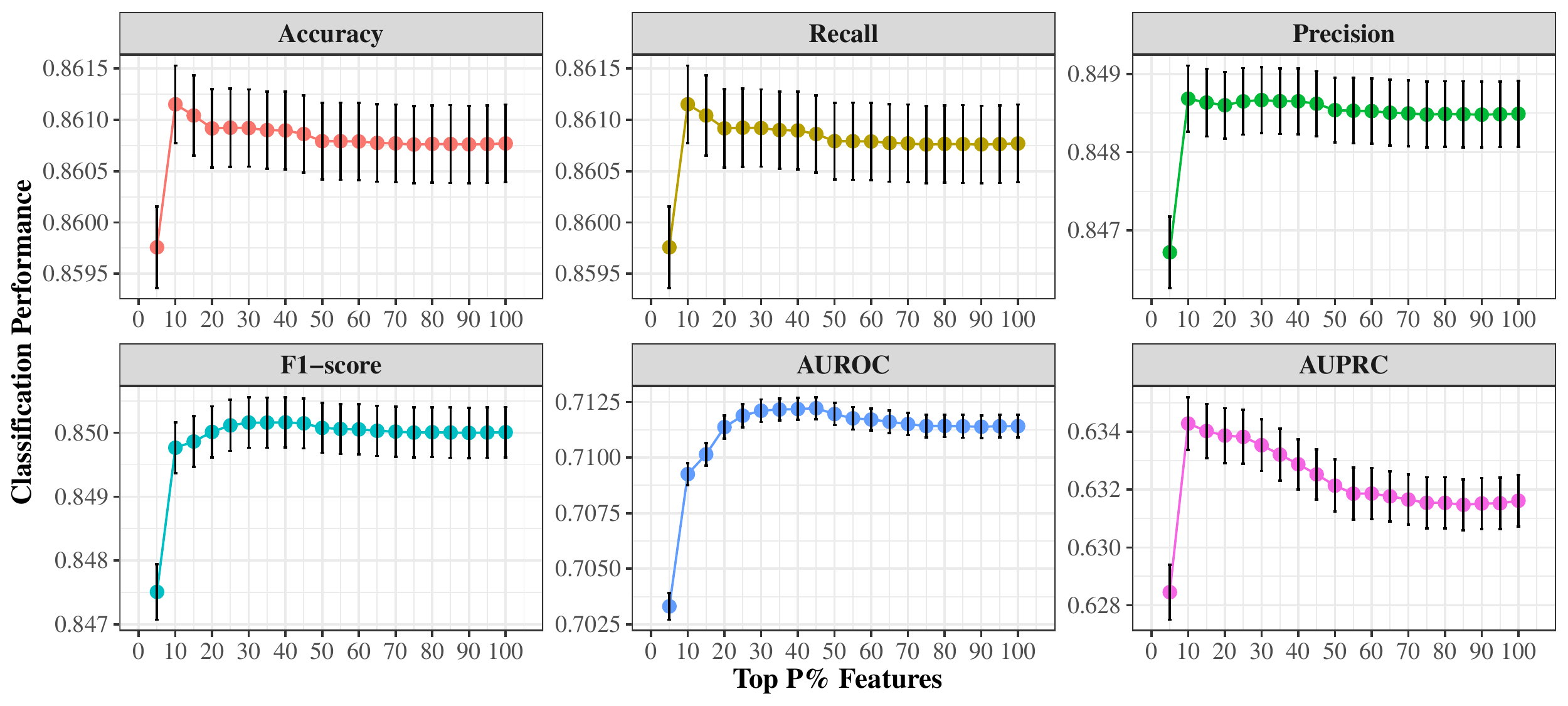}
\caption{DDI Classification Performance using Feature Subsets of Top $P\%$ Features (Mean $\pm$ 95\% CI, $N=30$)}
\label{top_percentile_perf}
\end{figure}

We test the classification performance of $F\textsubscript{P}$ $(P\in \{5,10,\cdots, 100\})$ over 30 repeats and calculate the mean and 95\% confidence interval (CI) of the evaluation metrics (Figure ~\ref{top_percentile_perf}). When the top 5\% features are used, all the metrics have the lowest value with statistical significance, which could be caused by insufficient number of features for classification. When the top 10\% features are used, accuracy, recall, precision and AUPRC reach the peak value. When $P$ increases from 10 to 100 with a step size 5, the metrics are either gradually decreasing or reached a plateau state. Overall, $F\textsubscript{10}$ can achieve comparable performance with $F\textsubscript{100}$, \textit{i.e.}, the whole feature set $F$. Therefore, $F\textsubscript{10}$ is the optimal feature subset.

\subsection*{3.2 Key Pharmacological Components in DDI}
\begin{figure}[h!]
\centering
\includegraphics[scale=0.54]{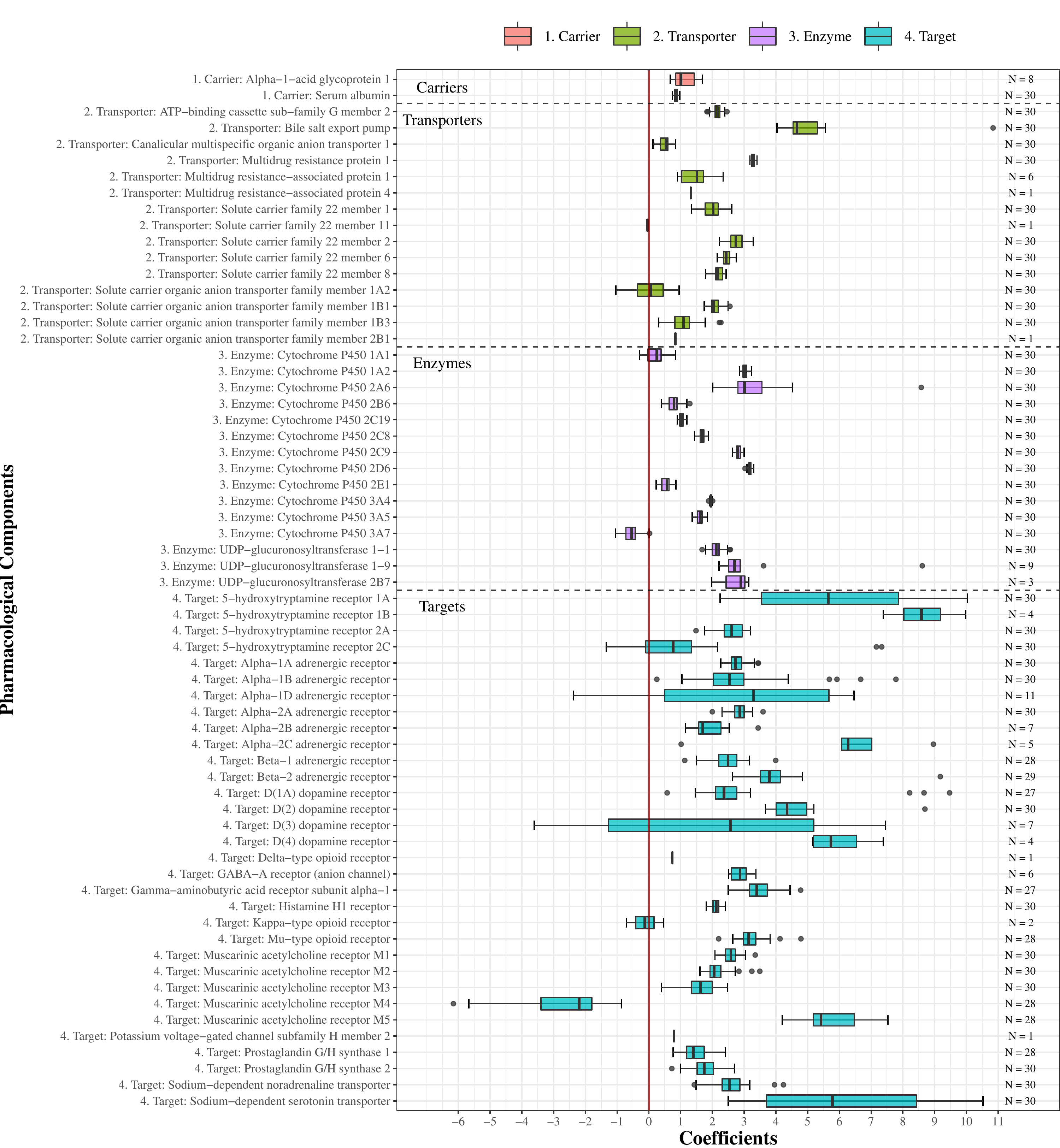}
\caption{Identified Important Pharmacological Components for DDI Classification}
\label{sel_ftrs_analyses}
\end{figure}

We extract individual features in all $F\textsubscript{10}$ over 30 repeats. To further compare their importance, we extract their corresponding $\beta$ coefficients in the LR classifier as well as the number of occurrences over 30 repeats (Figure ~\ref{sel_ftrs_analyses}). In total, there are 64 key pharmacological components identified as important in causing DDI, including 2 carriers (12.5\% of \textit{Carrier}), 15 transporters (23.1\% of \textit{Transporter}), 15 enzymes (24.2\% of \textit{Enzyme}) and 32 targets (9\% of \textit{Target}), which are discussed next. 

\textit{Key Carriers}

Serum albumin with a median $\beta$ coefficient around 1 is a plasma protein which can non-specifically bind drugs. Studies were conducted on its binding sites which can cause DDI \cite{zhu2008new}. Another carrier, alpha-1-acid glycoprotein 1, is also a carrier involved in the binding and transporting of multiple drugs. When two drugs can both bind to this carrier, either drug or both drugs can have altered distribution and a DDI is expected \cite{huang2013effect}.

\textit{Key Transporters}

Important transporters in DDI revealed by our study are mainly in two categories: ATP-binding cassette (ABC) transporters, such as ATP-binding cassette sub-family G member 2 (ABCG2, BCRP), bile salt export pump (BSEP) and Multidrug Resistance Protein 1 (ABCB1, MDR1, P-gp), and solute carrier family (SLC) transporters, such as solute carrier family 22 member 2 (SLC22A2, OCT2) and solute carrier organic anion transporter family 1B1 (SLCO1B1, OATP1B1). A systematic review on membrane transporters in drug development has pointed out that these transporters have a role in clinical DDI \cite{giacomini2010membrane}. For instance, P-gp distributed in intestine, kidney proximal tubule, liver and blood brain barrier can export a wide range of drugs, such as cyclosporine, doxorubicin, digoxin and loperamide. The role of P-gp in absorption, distribution and excretion makes it to have a high potential to cause DDI. Another transporter, BSEP, mainly located in hepatocytes and responsible for drug excretion, has median $\beta$ coefficient around 4.5, which indicates its prominent importance in DDI as well.

\textit{Key Enzymes}

Phase I metabolizing enzymes such as Cytochromes P450 (CYPs) have long been perceived as a key factor in DDI \cite{guengerich1997role}. It is reported that in liver, a major organ for drug metabolism, the highest expressed CYP isoforms include CYP1A2, CYP2C8, CYP2C9, CYP2E1 and CYP3A4, which are all manifested in the identification results \cite{zanger2013cytochrome}. Other isoform such as CYP2B6, which demonstrates clinical significance in recent years, is also captured \cite{zanger2013pharmacogenetics}. Several phase II metabolizing enzymes such as UDP-glucuronosyltransferases (UGT), including UGT1-1, UGT1-9 and UGT2B7, have median $\beta$ coefficients higher than 2, indicating a non-negligible role that they play in DDI. Unlike CYP mediating oxidation reactions, UGTs mainly catalyze conjugation reactions. For example, UGT2B7 converts codeine, an opioid anagesic, into codeine 6-glucuronide, a major active metabolite responsible for 60\% analgesic effect of codeine \cite{srinivasan1997analgesic}. Meanwhile, methadone, a medication to treat opioid use disorder, can inhibit UGT2B7 and substantially reduce the plasma concentration of codeine 6-glucuronide to form a DDI with codeine \cite{gelston2012methadone}.

\textit{Key Targets}

Identified targets mainly include the following categories: 1) 5-hydroxytryptamine (5-HT) receptors (\textit{a.k.a} serotonin receptors) and sodium-dependent serotonin transporters (\textit{a.k.a.}, serotonin transporter, SERT) which are the targets for many antidepressant medications, 2) adrenergic receptors which correspond to adrenergic drugs, 3) dopamine receptors whose targeted drugs are mainly used to treat Parkinson's disease, 4) opioid receptors which are the targets for opioid drugs, 5) GABA-A receptors whose agonists include benzodiazepines, 6) muscarinic acetylcholine receptors (\textit{a.k.a} cholinergic receptors) which correspond to antispasmodic drugs, and 7) prostaglandin G/H synthases (\textit{a.k.a.}, Cyclooxygenases, COX) which are the targets for non steroidal anti-inflammatory drugs (NSAIDs). 


Among the 64 identified pharmacological components, 50\% of them are targets. There are two reasons why drug targets consists of the major group. First, there is a large number of features in the \textit{Target} group. Secondly, PD DDI mediated by drug targets exist more widely than PK DDI, because target agonists or antagonists directly strengthen or weaken their therapeutic effect making PD DDI more likely to be detected. On the contrary, PK DDI mediated by carriers, transporters and enzymes can lead to concentration variations, which, however, may not necessarily exert a noticeable PD effect. 

\subsection*{3.3 Insights for DDI Study}
Two cases studies are conducted to generate insights for DDI studies. 

\textit{Trending Pharmacological Components Group in DDI}

\begin{figure}[h!]
\centering
\includegraphics[scale=0.6]{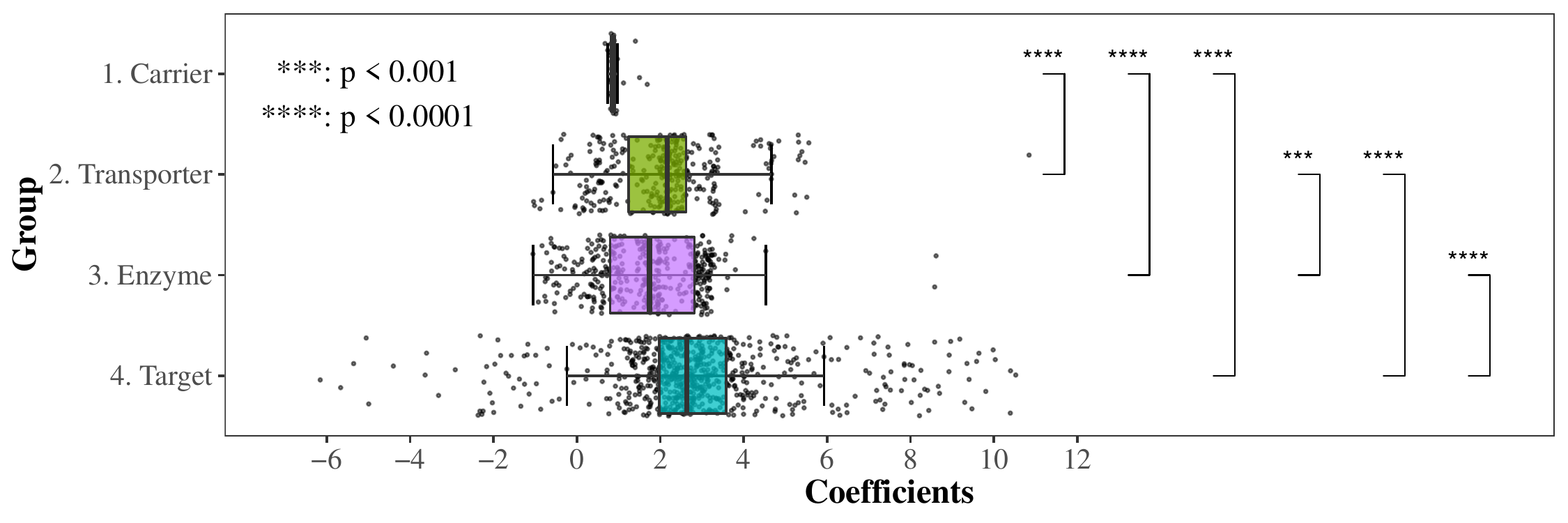}
\caption{Comparison of Coefficients in Different Pharmacological Components Groups}
\label{sel_ftrs_group_analyses}
\end{figure}

We aggregate all the identified pharmacological components by the four groups, i.e., \textit{Carrier}, \textit{Transporter}, \textit{Enzyme} and \textit{Target} (Figure~\ref{sel_ftrs_group_analyses}), where the unpaired two-samples t-test is used to compare whether there is statistically significant difference between different groups with regard to the $\beta$ coefficients. 

\textit{Target} is the most influential group in DDI, with $\beta$ coefficients significantly higher than the other three groups ($ p < 0.0001$). On the contrary, \textit{Carrier} is the least important group which has statistically lower coefficients than the other three groups ($ p < 0.0001$). For \textit{Enzyme} and \textit{Transporter}, \textit{Transporter} has slightly higher coefficients than \textit{Enzyme} with statistical significance ($ p < 0.001$), which demonstrates that drug transporters can contribute to DDI equally as enzymes. Interestingly, it is not until in recent years that the significance of transporters has been well addressed by researchers \cite{giacomini2010membrane}. Our findings can be corroborated by the current DDI study trends.

\textit{Novel Measure beyond Amount and Activity} 

\begin{figure}[h!]
\centering
\includegraphics[scale=0.7]{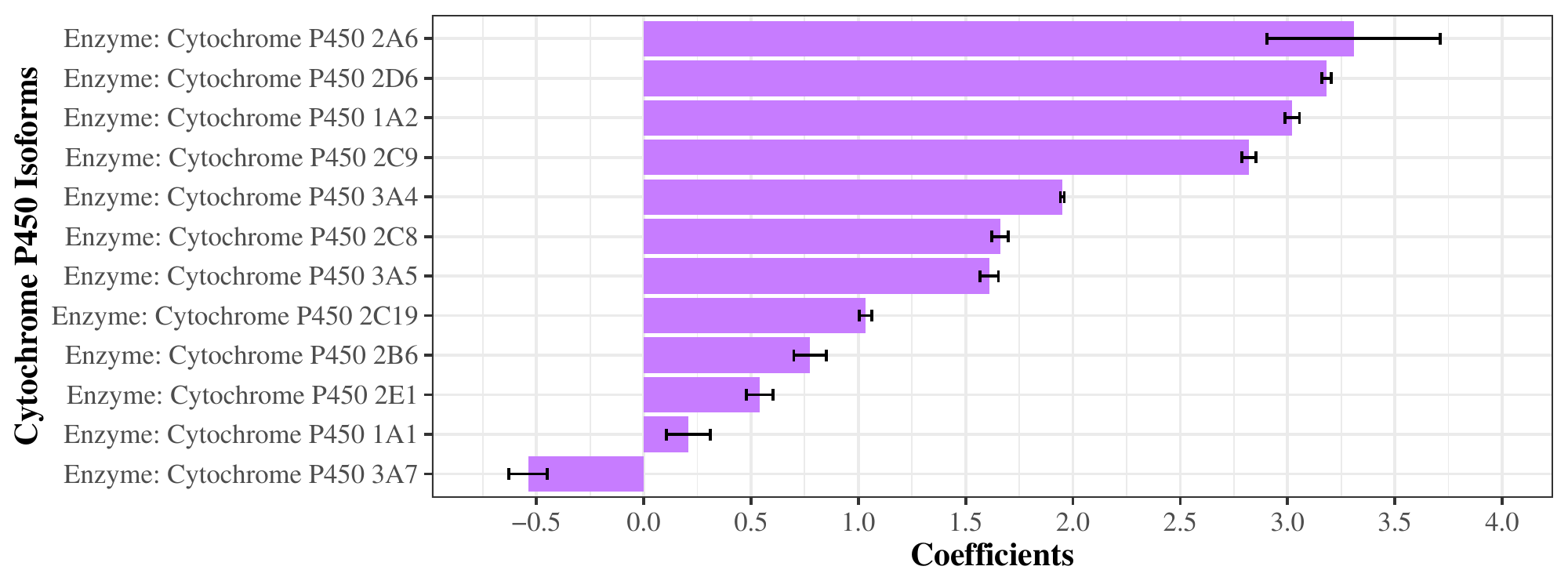}
\caption{Comparison of Different CYP Isoforms Coefficients (Mean $\pm$ 95\% CI, $N=30$)}
\label{top10_sel_ftrs_cyp_analyses}
\end{figure}

We extract all CYP isoforms and their corresponding coefficients over 30 repeats. Since all CYP isoforms occur in each repeat resulting in sample size at 30, the unpaired two-samples t-test is used to examine the statistical significance (Figure~\ref{top10_sel_ftrs_cyp_analyses}). CYP2A6, CYP2D6 and CYP1A2 with coefficients of no statistically significant differences are identified to be the most important CYP isoforms in DDI, closely followed by CYP2C9 $>$ CYP3A4 $>$ CYP2C8 $>$ CYP3A5 $>$ CYP2C19 $>$ CYP2B6 $>$ CYP2E1 $>$ CYP1A1 $>$ CYP3A7.

In traditional DDI studies, the amount and activity of CYP isoforms measured via \textit{in vitro} experiments provides prior knowledge to predict the importance of different CYP isoforms in drug metabolism and DDI \cite{zanger2013cytochrome}. In our study, feature importance can also act as a novel measure for DDI potential and allows evaluation of the CYP isoforms directly from a DDI-centered perspective. 

\section*{4. Discussion}
Traditional routine DDI studies such as wet-lab experiments often require broad knowledge on pharmacological components. Due to the high study cost and long experimental duration, it is infeasible to perform experiments on all pharmacological components. Identification of most relevant carriers, transporters, enzymes and targets is critical in wet-lab studies. 

Different from other informatics-based DDI studies, our study focus on utilizing large-scale information on DDI to generate insights and boost wet-lab study efficiency. In order to answer the question ``which pharmacological components are most relevant to DDI'', we conduct an informatics-based DDI study to demystify the underlying key carriers, enzymes, targets and transporters in DDI. A machine-learning methodology based on classification and feature selection is developed to identify key pharmacological components and strong literature evidence supports our findings. Furthermore, by exploiting logistic regression, a classifier with interpretable feature coefficients, the importance of the key components is also quantified, which can act as a novel and direct DDI-oriented measure to evaluate the pharmacological components.

Our approach is not without its limitations. First, the nature of interaction between the drug and the pharmacological component is not considered.  At this stage, we only take into account whether a drug interacts with a specific pharmacological component. In the current feature engineering process, we do not capture how a drug induces or inhibits the pharmacological component, or if the drug is merely a substrate. The relationship between a drug and its corresponding pharmacological component will be further explored. Secondly, DrugBank may not have the most complete DDI information, and we will explore other potential data sources.  In the future, in addition to the logistic regression classifier, we will also explore deep learning classifiers, in particular those with interpretable capability. 

\section*{5. Conclusion}
In this paper, we use an informatics-based approach to study which pharmacological components are most relevant to DDI. We identify a list of key pharmacological components including 2 carriers, 15 transporters, 15 enzymes and 32 targets. The quantified importance of these components in DDI in our study is consistent with current DDI study trends and acts as a novel measure for pharmacological components from a DDI-oriented perspective. 

\makeatletter
\renewcommand{\@biblabel}[1]{\hfill #1.}
\makeatother


\bibliographystyle{unsrt}
\bibliography{reference}


\end{document}